\documentclass[aps,prd,showpacs,preprintnumbers,nofootinbib]{revtex4}
\usepackage[dvips]{graphicx}
\usepackage{bm,latexsym,amsmath,amssymb,amsfonts}

\usepackage[normalem]{ulem}

\begin{document}

\title{\uline{}Gravitational collapse of charged dust cloud in the Lovelock gravity}

\author{Seiju Ohashi, Tetsuya Shiromizu}
\affiliation{Department of Physics, Kyoto University, Kyoto 606-8502, Japan}

\author{Sanjay Jhingan}
\affiliation{Centre for Theoretical Physics, Jamia Millia Islamia, New Delhi 110025, India}

\begin{abstract}
We study the effect of charge on gravitational collapse of inhomogeneous dust 
cloud in the Einstein, Gauss-Bonnet and Lovelock gravity. Dynamics of the collapsing 
shell is analyzed. The conditions for the occurrence of bounce during collapse are given. 
We also show that shell crossing occurs inevitably if the shell near the center is weakly charged. 
\end{abstract}
\maketitle

\section{Introduction}

The final fate of gravitational collapse is one of the important unsolved problems in 
classical general relativity (GR)\textsl{}. Both the possibilities, i.e., the formation 
of naked singularity or black hole as an end-state of collapse, may present a  crucial 
step in our understanding of nature. On one hand, we have the singularity theorems which 
tell us that singularities are inevitably formed through gravitational collapse with 
physically reasonable and general initial data \cite{Hawking:1973uf}. These singularities 
lead to a loss of predictability if visible to a distant observer, and hence a 
break down of GR happens. On the other hand, the formation of naked singularities  can imply an 
access to strong gravity regions. That is, they offer a possibility to probe the laws 
of nature and a window to a quantum gravity theory. However, one may wish that 
singularities should be wrapped by event horizons. This idea is paraphrased in the form 
of ``cosmic censorship conjecture" (CCC) \cite{Penrose:1969pc}. If this conjecture is 
indeed true, we have black holes. A proof or otherwise of the CCC remains one of the 
unresolved issues in classical GR. 

The first pioneering work in gravitational collapse was carried out by Oppenheimer and Snyder 
\cite{Oppenheimer:1939ue}. They considered the spherical collapse of a homogeneous dust cloud.
They showed that black hole is formed and the singularity is enclosed by the event horizon.
However, many solutions which contain naked singularities have been found, for example, 
in the collapse of inhomogeneous dust cloud 
\cite{Yodzis:1973,Eardley:1978tr,Christodoulou:1984mz,Newman:1985gt,Joshi:1993zg,Singh:1994tb,Jhingan:1996jb,Joshi:2008zz}.
We now understand that, in the Oppenheimer-Snyder model, the homogeneity of dust cloud and 
spherical symmetry are important for the formation of black holes. 
In order to examine the generality of the above results, it is important to study the collapse 
with generic initial conditions. Then, gravitational collapse has been studied 
in various situations, {\it e.g.}, collapse of charged dust cloud, dust with tangential pressure, 
perfect fluid and so on \cite{Krasinski:2006sb,Harada:1998cq,Jhingan:1999ht,Goncalves:2001eb,Jhingan:2002,Malafarina:2010xs}. 

In the late stages of collapse with the strong gravity, GR does not provide us with an accurate 
description of the final fate of gravitational collapse. Motivated by this and the fact that the 
best candidate fundamental theory 
such as  superstring/M theory lives naturally in more than four  dimensions, considerable 
efforts have been made on studies of gravitational collapse in higher dimensional gravitational 
theories. Gravitational collapse in the simple extensions of GR to higher dimension is studied 
as well \cite{Ghosh:2001fb,Goswami:2004gy,Goswami:2006ph}. It has been found that the singularity 
formed through the spherical collapse of inhomogeneous dust cloud cannot be naked in dimensions higher than five with smooth initial data. However, spacetime around the singularity is extremely 
curved and the reliability of the pure GR is doubtful. The physical motivation to go beyond 
GR comes from the fact that quantum gravitational effects/string corrections play a significant role in the
region near singularity. Actually, superstring theory predicts certain higher 
order curvature corrections to GR. From this fact it is important to study the gravitational 
collapse in theories with higher curvature corrections. 

 In this paper we focus on a special type of gravitational theory with higher curvature 
corrections, that is the Lovelock theory \cite{Lovelock:1971yv}. 
The Lovelock theory is the most general gravitational theory in which: (a) Lagrangian 
is covariant and (b) field equations contain up to second order derivatives of metric tensor.
The Lovelock gravity is a widely studied higher dimensional gravitational theory. The black 
hole solutions were found in Refs \cite{Boulware:1985wk, Wheeler:1985qd, Cai:2003kt, Cai:2001dz}. 
Some authors have studied the dust collapse in a special case of the Lovelock theory \cite{Maeda:2006pm,Nozawa:2005uy}. 
In our previous paper \cite{Ohashi:2011zz}, we studied the spherical collapse of dust cloud in the 
full Lovelock theory with any spacetime dimension. We showed that the singularities are always formed and can be 
naked for some initial data. In addition we showed that the nature of singularity 
depends on the dimensionality of spacetime, that is, whether it is even or odd. 

As mentioned before, it is important to investigate the collapse in more general situations 
from the CCC point of view. To this end we will consider the spherical collapse of a dust cloud 
coupled with the Maxwell field, {\it i.e.}, a charged dust cloud, in the Einstein, Gauss-Bonnet 
and Lovelock gravity. The collapse of a charged dust cloud can be regarded as a toy 
model for understanding rotating spacetimes. We will also compare the dynamics of a charged dust cloud with a neutral 
dust cloud. In the charged dust case, using a kinematical argument, we see that 
the shell near the center inevitably bounces if the charge is sufficiently small. We also show that the 
shell-crossing singularities occurs 
before the shell bounces. This feature is quite different from the neutral dust cloud cases \cite{Ohashi:2011zz}. 
Although we have included higher order curvature corrections, our result is same as in the case of the 
Einstein gravity(See Refs. \cite{Ori:1990, Ori:1991}.). This is quite impressive because the higher 
curvature corrections do affect the dynamics in the case of neutral dust cloud. 

The rest of this paper is organized as follows. In Sec. II, we briefly review the Lovelock theory. 
In Sec. III, we derive the basic equations and solve these partially to discuss the shell motion kinematically. 
Looking at the key equations, in Sec. IV, we show that the shell-bounce will occur kinematically. 
We investigate the Einstein, Gauss-Bonnet and full Lovelock gravity case in turn. In Sec. V, 
we show that shell-crossings inevitably occur before the occurrence of the shell-bounce. In Sec. VI, 
we briefly summarize and discuss our result. In the appendices, we consider the matching condition of 
the inner collapsing with the outer electro-vacuum region and a detail of the calculation in the text. 
We use the geometrized units, $G=c=1$. 

\section{lovelock gravity: the Exterior geometry}

In this section, we briefly review the Lovelock theory of gravity in $D=(n+2)$-dimensional spacetimes. 
The Lagrangian of the theory is
\begin{equation}
\mathcal{L}=\sum_{m=1}^{k}\frac{a_m}{m}\mathcal{L}_m ,
\end{equation}
where
\begin{equation}
\mathcal{L}_{m}=\frac{1}{2^m}\delta_{\mu_1 \mu_2 \dots \mu_{2m-1} \mu_{2m}}^{\nu_1 \nu_2 \dots \nu_{2m-1} \nu_{2m}}R_{\nu_1 \nu_2}{}^{\mu_1 \mu_2 }\dots R_{\nu_{2m-1}\nu_{2m}}{}^{\mu_{2m-1}\mu_{2m}}{} .
\end{equation}
%
Here $R_{pq}{}^{sr}$ is the Riemann curvature tensor of spacetime and 
$\delta_{\mu_1 \mu_2 \dots \mu_{2m-1} \mu_{2m}}^{\nu_1 \nu_2 \dots \nu_{2m-1} \nu_{2m}}$ is 
the generalized totally antisymmetric Kronecker delta.  $a_m$ are arbitrary constants which 
cannot be determined by the theory. The suffixes $\mu_1, \dots ,\mu_{2m}$ and $\nu_1, \dots ,\nu_{2m}$ 
run from $1$ to $D$, and $k$ is given by 
$k=[(D-1)/2]$ ($[x]$ is the integer part of $x$).

The field equations derived from this Lagrangian are of the form
%
\begin{align}
\mathcal{G}_{\mu }^{\nu }&=-\sum_{m=1}^{k}\frac{1}{2^{m+1}}\frac{a_m}{m}\delta_{\mu \mu_1 \mu_2 \dots \mu_{2m-1} \mu_{2m}}^{\nu \nu_1 \nu_2 \dots \nu_{2m-1} \nu_{2m}}R_{\nu_1 \nu_2}{}^{\mu_1 \mu_2 }\dots R_{\nu_{2m-1}\nu_{2m}}{}^{\mu_{2m-1}\mu_{2m}} \notag \\
&=8\pi T_{\mu}^{\nu} . \label{FEQ}
\end{align}
%
Here $T^{\nu}_{\mu}$ is the energy-momentum tensor of matter fields.

The static black hole solution with spherically symmetry was found in Refs. 
\cite{Wheeler:1985qd,Cai:2003kt,Cai:2001dz} for vacuum and in Refs. \cite{Charmousis:2008kc,Garraffo:2008hu} 
for electro-vacuum cases. An electro-vacuum spacetime is described by the metric of the form 
%
\begin{equation}
ds^2=-f(r)dt^2+\frac{1}{f(r)}dr^2+r^2\gamma_{ij}dx^idx^j .
\end{equation}
%
Here $\gamma_{ij}$ is the line element of $n$-dimensional maximally symmetric surfaces and $f$ is
%
\begin{equation}
f(r)=\kappa -r^2\psi (r) .
\end{equation}
%
The curvature constant $\kappa$ can take values $-1, 0$ and $1$.  The function $ \psi (r)$ is a 
solution of the algebraic equation
%
\begin{equation}
\sum_{m=2}^{k} \bigg[ \frac{a_m}{m}\bigg\{ \prod_{p=1}^{2m-2}(n-p)\bigg\} \psi^m \bigg] +\psi =\frac{\mu}{r^{n+1}}-\frac{q^2}{r^{2n}},
\end{equation}
%
where $\mu$ and $q$ are constants proportional to the Arnowitt-Deser-Misner (ADM) mass and charge, 
respectively. This solution describes the spacetime exterior to a charged dust cloud (See Appendix A 
for the details). In what follows we shall restrict ourselves to the case  $\kappa=1$ because we are interested in 
asymptotically flat spacetimes here. The generalization of our study to other cases with $\kappa=0,-1$ 
is rather easy. 
 
\section{Basic Equations: the Interior geometry}

In this section we shall derive equations for the interior of a collapsing charged dust cloud. 
The spacetime metric under  the  spherical symmetry ansatz can be written as
%
\begin{equation}
ds^2=-A^2dt^2+B^2 dr^2+R^2 d\Omega^2_n \label{sphmetric},
\end{equation}
%
where the metric functions  $A(t,r), B(t,r)$ and $R(t,r)$ are arbitrary functions of coordinates 
$t$ and $r$, and $d\Omega^2_n$ is the line element of a $n$-sphere. Using the gauge freedom 
we set $R(0,r)=r$ on an initial data surface.  We also assume that the interior of the collapsing 
cloud is composed of  charged dust, {\it i.e.} the energy momentum tensor $T^{\mu}{}_{\nu}$ takes a form
%
\begin{equation}
T^{\mu}{}_{\nu}=T^{(D)}{}^{\mu}{}_{\nu}+T^{(M)}{}^{\mu}{}_{\nu}\;.
\end{equation}
%
Here $T^{(D)}{}^{\mu}{}_{\nu}$ is the energy momentum tensor for a dust cloud
%
\begin{equation}
T^{(D)}{}^{\mu}{}_{\nu}=\varepsilon(t,r) u^{\mu}u_{\nu}
\;,\end{equation}
and $T^{(M)}{}^{\mu}{}_{\nu}$ is that of the Maxwell field,\begin{equation}
T^{(M)}{}^{\mu}{}_{\nu}=\frac{1}{4\pi} \left( F^{\mu \rho}F_{\nu \rho}-\frac{1}{4}g^{\mu}{}_{\nu}F^{\rho \lambda}F_{\rho \lambda} \right) .
\end{equation}
%
$\varepsilon(t,r) $ is the energy density of the dust cloud, $u^{\mu}$ is its 4-velocity 
and $F_{\mu \nu}$ is the field strength of the Maxwell field. In addition, we have the 
Maxwell's equations  given by
%
\begin{equation}
\nabla_{\mu}F^{\mu \nu}=-4\pi j^{\nu}\;,
\end{equation}
and
\begin{equation}
\nabla_{[\mu}F_{\nu \rho ]}=0 \;.
\end{equation}
%
Here $j^{\nu}$ is the electric current and $\nabla_{\mu}$ is the covariant derivative 
with respect to the spacetime metric. We assume that the current $j^{\mu}$ has the form
%
\begin{equation}
j^{\nu}=\sigma u^{\nu} \;,
\end{equation}
%
where $\sigma$ is the electric charge density.

In the case of a non-zero current, the Maxwell's equations become
%
\begin{equation}
\frac{1}{ABR^n}\partial_r \left( ABR^nF^{tr}\right) =\frac{4\pi}{A} \sigma
\end{equation}
and
\begin{equation}
\frac{1}{ABR^n}\partial_t \left( ABR^nF^{tr}\right) =0 .
\end{equation}
%
We can solve these equations as
%
\begin{equation}
F^{tr}=\frac{n}{2}\frac{Q(r)}{ABR^n} \label{fieldstr},
\end{equation}
%
where $Q(r)$ is the integration function and it is 
the charge contained in the ball of radius $r$, and it satisfies
%
\begin{equation}
Q^{\prime}=\frac{2}{n}4\pi \sigma BR^n \label{mxcharge}.
\end{equation}
%
The prime stands for the differentiation with respect to $r$. 

The conservation equations of the energy momentum tensor, $\nabla_{\nu}T^{\nu}_{\mu}=0$, lead to
%
\begin{equation}
\nabla_{\mu}\left( \varepsilon u^{\mu}\right) =0
\end{equation}
and
\begin{equation}
\varepsilon u^{\nu}\nabla_{\nu}u^{\mu}=\sigma F^{\mu}{}_{\nu}u^{\nu} .\label{modgeo}
\end{equation}
%
Using Eqs. (\ref{sphmetric}), (\ref{fieldstr}) and (\ref{mxcharge}), the
conservation equations  become
\begin{equation}
 \partial_t \left( \varepsilon  BR^n\right) =0  \label{conserve1}
\end{equation}
and
\begin{equation}
 \varepsilon\left( \frac{A^{\prime}}{A}\right) =\left( \frac{n}{2}\right)^2 \frac{QQ^{\prime}}{4\pi R^{2n}}  .\label{conserve2}
\end{equation}
%
Eq. (\ref{conserve1}) can be integrated as
%
\begin{equation}
N^{\prime}(r)=\frac{2}{n}4\pi \varepsilon BR^n ,\label{defN}
\end{equation}
where $N(r)$ is an arbitrary function of $r$. Using Eq. (\ref{defN}), we can re-express Eq. (\ref{conserve2}) as
\begin{equation}
\frac{A^{\prime}}{A}=\frac{n}{2}\frac{QQ_{N}B}{R^n} ,\label{eqA}
\end{equation}
where
\begin{equation}
Q_{N}\equiv \frac{Q^{\prime}}{N^{\prime}} =\frac{\sigma}{\varepsilon}.\label{QNse}
\end{equation}

Now, let us consider the field equations (\ref{FEQ}). They have the following nontrivial components:
%
\begin{align}
{}&\sum_{m=1}^{k}a_m\prod_{p=0}^{2m-2}(n-p)\left(\left( \frac{\dot{R}}{AR}\right)^2-\left( \frac{R^{\prime}}{BR}\right)^2  +\frac{1}{R^2}\right)^{m-1}\notag \\
&\ \ \ \ \ \ \ \ \ \times \left( \frac{\ddot{R}}{A^2R}-\frac{\dot{A}\dot{R}}{A^3R}-\frac{A^{\prime}R^{\prime}}{AB^2R} +\frac{\left( n-\left( 2m-1\right) \right)}{2m}\left( \left( \frac{\dot{R}}{AR}\right)^2-\left( \frac{R^{\prime}}{BR}\right)^2  +\frac{1}{R^2}\right) \right)
                   =\left( \frac{n}{2}\right)^2 \frac{Q^2}{R^{2n}} \label{LL1} ,\\ \notag \\
{}&\sum_{m=1}^{k}a_m\prod_{p=0}^{2m-2}(n-p)\left( \left( \frac{\dot{R}}{AR}\right)^2-\left( \frac{R^{\prime}}{BR}\right)^2  +\frac{1}{R^2}\right)^{m-1} \notag \\
&\ \ \ \ \ \ \ \ \ \times \left( -\frac{R^{\prime \prime }}{B^2R}+\frac{B^{\prime}R^{\prime}}{B^3R} +\frac{\dot{B}\dot{R}}{A^2BR} +\frac{\left( n-\left( 2m-1\right) \right) }{2m} \left( \left( \frac{\dot{R}}{AR}\right)^2-\left( \frac{R^{\prime}}{BR}\right)^2  +\frac{1}{R^2}\right) \right)
                   =8\pi \varepsilon (t,r)+\left( \frac{n}{2}\right)^2 \frac{Q^2}{R^{2n}} \label{LL2} \\ \notag
\end{align}
and
\begin{equation}
\sum_{m=1}^{k}a_m\prod_{p=0}^{2m-2}(n-p)\left( \left( \frac{\dot{R}}{AR}\right)^2-\left( \frac{R^{\prime}}{BR}\right)^2  +\frac{1}{R^2}\right)^{m-1}  \left( \frac{\dot{R}^{\prime}}{B^2R}-\frac{A^{\prime}\dot{R}}{AB^2R}-\frac{\dot{B}R^{\prime}}{B^3R} \right)
                  =0 \label{LL3} .
\end{equation}
%
In the above, the dot stands for the differentiation with respect to $t$. 

First we realize that, for a non-trivial solution, Eq. (\ref{LL3}) implies
%
\begin{equation}
\frac{\dot{R}^{\prime}}{B^2R}-\frac{A^{\prime}\dot{R}}{AB^2R}-\frac{\dot{B}R^{\prime}}{B^3R} =0 \label{solgtr}.
\end{equation}
%
Using Eq. (\ref{eqA}), we can easily integrate Eq. (\ref{solgtr}) as
%
\begin{equation}
\frac{R^{\prime}}{B}=W(r)-\frac{nQQ_{N}}{2(n-1)R^{n-1}} \label{A},
\end{equation}
%
where $W(r)$ is the integration function which corresponds to the total energy, as we will show later. 

Now we introduce a useful function $L$ defined as 
%
\begin{equation}
L\equiv\sum_{m=1}^k\frac{a_m}{2m}\prod_{p=0}^{2m-2}(n-p)\left( \left( \frac{\dot{R}}{A}\right)^2 +1-\left( W-\frac{nQQ_{N}}{2(n-1)R^{n-1}}\right)^2 \right)^{m}\frac{1}{R^{2m}} \label{defL}.
\end{equation}
%
Using this definition of $L$, we can rewrite Eqs. (\ref{LL1}) and (\ref{LL2}) in a simple form, i.e.
%
\begin{equation}
\frac{1}{\dot{R}R^n}\frac{\partial}{\partial t}\left( LR^{n+1}\right) =\left( \frac{n}{2}\right)^2 \frac{Q^2}{R^{2n}} \label{simpleLL1}
\end{equation}
and
\begin{equation}
\frac{1}{R^{\prime}R^n}\frac{\partial}{\partial r}\left( LR^{n+1}\right) =8\pi \varepsilon + \left( \frac{n}{2}\right)^2\frac{Q^2}{R^{2n}}\label{simpleLL2}.
\end{equation}
%
The integration of Eq. (\ref{simpleLL1}) gives us
%
\begin{equation}
L=-\left( \frac{n}{2}\right)^2\frac{Q^2}{(n-1)R^{2n}}+\frac{n}{2}\frac{F(r)}{R^{n+1}} \label{LL6} ,
\end{equation}
%
where $F(r)$ is the integration function which corresponds to the mass, as shown later. Substituting this into Eq. (\ref{simpleLL2}), we see that 
%
\begin{equation}
8\pi \varepsilon =\frac{n}{2}\frac{F(r)^{\prime}}{R^nR^{\prime}}-\frac{n^2QQ^{\prime}}{2(n-1)R^{2n-1}R^{\prime}} .\label{simpleLL3}
\end{equation}
%
From this expression, we see that the spacetime admits two type of singularities in general. The 
shell-focusing singularity at $R=0$ and the shell-crossing singularity at $R^{\prime}=0$.
Using Eq. (\ref{A}), we can re-express mass the function $F(r)$ as
%
\begin{equation}
F(r)^{\prime}=2WN(r)^{\prime}. \label{FW}
\end{equation}
%

Combining Eq. (\ref{defL}) with Eq. (\ref{LL6}), we can obtain the basic equation which determines $R$
%
\begin{equation}
\sum_{m=1}^kc_m\left( \left( \frac{\dot{R}}{A}\right)^2 +1-\left( W-\frac{nQQ_{N}}{2(n-1)R^{n-1}}\right)^2 \right)^{m}\frac{1}{R^{2m}}=-\left( \frac{n}{2}\right) \frac{Q^2}{(n-1)R^{2n}}+\frac{F(r)}{R^{n+1}} \label{basiceq},
\end{equation}
%
where $c_l$ is defined by 
%
\begin{equation}
c_l =\begin{cases}
     1 & \text{if}\ \  l=1\\
     \frac{a_l}{l}\prod_{p=1}^{2l-2}(n-p) & \text{if}\ \  2\leq l \leq k .
     \end{cases}
\end{equation}
%
In order for the initial surface to be regular at the center $(r=0)$, the functions 
$F(r),Q(r)$ and $W(r)$ should behave as
%
\begin{align}
F(r)&=r^{n+1}M(r) \label{regularity}, \\
Q(r)&=r^{n+1}q(r) \label{regularity2}
\end{align}
and
\begin{equation}
W(r)=1+r^2b(r),
\end{equation}
%
where $M(r),q(r)$ and $b(r)$ are regular functions on the initial hypersurface, and it is supposed 
that $q(r)$ and $M(r)$ behave as $q'(r),M'(r)=\mathcal{O}(r)$ near $r=0$.

Before closing this section, we would like to make a remark.  
One might wonder whether
\begin{equation}
\sum_{m=1}^{k}a_m \prod_{p=0}^{2m-2}(n-p)\left( \left( \frac{\dot{R}}{RA}\right)^2
-\left( \frac{R^{\prime}}{RB}\right)^2  +\frac{1}{R^2}\right)^{m-1} =0
\end{equation}
is also a solution to Eq. (\ref{LL3}).
But, combining with Eq. (\ref{LL1}), we have
\begin{eqnarray}
\sum_{m=1}^{k}a_m\prod_{p=0}^{2m-2}(n-p)\frac{\left( n-\left( 2m-1\right) \right)}{2m}
 \left(\left( \frac{\dot{R}}{RA}\right)^2
-\left( \frac{R^{\prime}}{RB}\right)^2  +\frac{1}{R^2}\right)^{m}
=\left( \frac{n}{2}\right)^2 \frac{Q^2}{R^{2n}}.\label{another}
\end{eqnarray}
We can see that Eqs. (\ref{LL2}) and (\ref{another}) imply $\varepsilon (t,r)=0$.
Hence, the spacetime must be electro-vacuum which is of no interest to us here.

\section{Analysis of Shell Motion}

In this section we shall study a weakly charged shell in the vicinity of the center ($r=0$)
and show that its bounce is inevitable. The analysis in this section is also used to 
show the occurrence of shell-crossing singularity in the next section. For a pedagogical 
reason, we shall start with  the Einstein gravity case and then discuss the 
Gauss-Bonnet and full Lovelock gravity cases.

\subsection{Einstein gravity case}

Here we investigate the spherical collapse of charged dust in the Einstein gravity 
with arbitrary dimensions. In this case, Eq. (\ref{basiceq}) becomes
%
\begin{equation}
\left( \frac{\dot{R}}{A}\right)^2 =-V_{E}, \label{dRVE}
\end{equation}
where
\begin{equation}
V_E \equiv -E(r) - \frac{M(r)}{R^{n-1}}
- \left( \frac{n}{2}\right)^2 \frac{Q^2}{(n-1)^2} \left( Q^2_{N}-\frac{2(n-1)}{n}\right)\frac{1}{R^{2(n-1)}}\;. 
\end{equation}
In the above $E(r) \equiv W^2(r)-1$ and  $M(r) \equiv F(r)-{nWQQ_N}/{(n-1)} $.

Let us consider whether  shell bounce ($V_{E}=0$) occurs or not. From the discriminant of 
$V_E=0$ with respect to $R$, we find that there is no solution of $R$ for $V_{E}=0$ if
%
\begin{equation}
M^2(r)< \frac{4E(r)Q^2(r)}{(n-1)^2}\left( Q^2_N-\frac{2(n-1)}{n}\right) \label{nosol}.
\end{equation}
%
Thus, if Eq. (\ref{nosol}) is satisfied, (i) shell always collapses into shell-focusing singularity($R=0$) 
because $V_E$ stays negative definite throughout collapse
in the $E\geq 0$ case, and (ii) there is no solution for Eq. (\ref{dRVE}) in the $E<0$ cases 
because $V_E$ always has a positive value.

Similarly we can analyze different cases for the condition
%
\begin{equation}
M^2(r)\geq \frac{4E(r)Q^2(r)}{(n-1)^2}\left( Q^2_N-\frac{2(n-1)}{n}\right) \label{solex}.
\end{equation}
%
In this case there are various possible situations depending on the sign of $E,M$ and $Q^2_N-{2(1-1/n)}$. For example, in the case for $E>0, M>0$ and $Q^2_N-2(1-1/n)>0$, $V_E$ always has negative value. This implies that the shell can hit the shell focusing singularity. On the other hand, in the case for $E>0, M>0$ and $Q^2_N-2(1-1/n)<0$, $V_E$ cannot have negative value for small $R$. This means shell inevitably bounces. 
In Table $1$, we summarize the result satisfying the condition above. 

\begin{center}
\begin{tabular}{|c|c|c|c|c|c|c|} \toprule
 &\multicolumn{2}{c|}{$E>0$}&\multicolumn{2}{c|}{$E=0$}&\multicolumn{2}{c|}{$E<0$} \\ \hline
 &$M>0$&$M<0$&$M>0$&$M<0$&$M>0$&$M<0$ \\ \hline
$Q^2_N>\frac{2(n-1)}{n}$& singular& {singular and bounce} & singular &singular &singular &singular \\ \hline
$Q^2_N<\frac{2(n-1)}{n}$& bounce &bounce &bounce &no solution & {oscillations} &no solution \\ \toprule
\end{tabular}\\
\vspace{0.1in}
{Table $1$ The cases satisfying condition Eq. (\ref{solex}).}
\end{center}

In summary we could show that (i) every shell with $Q^2_N< {2(1-1/n)}$ bounces if the solution to 
Eq. (\ref{dRVE}) exists, and (ii) every shell with $Q^2_N>{2(1-1/n)}$ collapses into shell-focusing singularity.
In the Einstein gravity, the condition, $Q^2_N<{2(1-1/n)}$ seems to play a significant role for occurrence of the bounce. From Eq. (\ref{defN}), 
we can express this condition as
%
\begin{equation}
|\sigma |< \left[ {2\left(1-\frac{1}{n}\right)}\right]^{1/2} \varepsilon \label{eincondi}.
\end{equation}
%
This means that a weakly charged collapsing shell will inevitably bounce in the case of 
the Einstein gravity. The condition for the bounce depends on the ratio of the charge to the dust
density and the number of spacetime dimensions. Note that this result is basically same as that obtained for 
four dimensions \cite{Ori:1990, Ori:1991}.

\subsection{Gauss-Bonnet/Lovelock gravity}

Next we shall consider the full Lovelock gravity cases including the Gauss-Bonnet one. 
Eq. (\ref{basiceq}) becomes
%
\begin{equation}
\sum_{m=1}^kc_m\left( \left( \frac{\dot{R}}{A}\right)^2 +1-\left( W-\frac{nQQ_{N}}{2(n-1)R^{n-1}}\right)^2 \right)^{m}\frac{1}{R^{2m}}=-\left( \frac{n}{2}\right)\frac{Q^2}{(n-1)R^{2n}}+\frac{F(r)}{R^{n+1}} 
\label{EQGB}.
\end{equation}
%

Let us consider a certain epoch $t=t_*$ when
%
\begin{equation}
R^{n-1}=\frac{nQ^2}{2(n-1)F}
\end{equation}
%
holds.  At this epoch, the right-hand side of Eq. (\ref{EQGB}) vanishes. To achieve this condition in 
collapsing situations, the condition $R^{n-1} \geq \frac{nQ^2}{2(n-1)F}$ 
should be satisfied at the initial surface. We can see that this is always satisfied near the center($r=0$) 
because the initial condition for $R$ is $R=r$ and the regularity condition for $F(r)$ and $Q(r)$ 
implies that $\frac{nQ^2}{2(n-1)F} $ goes as $\mathcal{O}(r^{n+1})$ near the center.

At $t=t_*$ we can rewrite  Eq. (\ref{EQGB}) as
%
\begin{equation}
\sum_{m=1}^kc_m\left( \left( \frac{\dot{R}}{A}\right)^2 +1-\left( W-\frac{FQ_{N}}{Q}\right)^2 \right)^{m}\frac{1}{R^{2m}}=0.
\end{equation}
%
Here we focus on the weakly charged cases which satisfy the condition
%
\begin{equation}
\frac{W-1}{2W}\frac{F^{\prime}}{F}\leq \frac{Q^{\prime}}{Q} \leq \frac{W+1}{2W}\frac{F^{\prime}}{F}\;, \label{awkcondi2}
\end{equation}
%
and one can show
%
\begin{equation}
\left( \frac{\dot{R}}{A}\right)^2 \leq 0 .
\end{equation}
But this is impossible. Therefore we may conclude that the shells near the center will inevitably bounce at 
least before $t=t_*$, if the artificial condition (\ref{awkcondi2}) is satisfied. 

Next, let us briefly examine the physical interpretation of the condition (\ref{awkcondi2}). 
In the vicinity of the center, $W(r)$ behaves as $W(r)=1+b(r)r^2$ and then the condition (\ref{awkcondi2}) 
can be written approximately as
%
\begin{equation}
0\lesssim \frac{Q^{\prime}}{Q} \lesssim \frac{F^{\prime}}{F}.
\end{equation}
%
Roughly speaking, this is satisfied when the spatial scale of the mass distribution is shorter than charge one.
The integration of the above gives us
%
\begin{equation}
\frac{Q(r)}{F(r)}\lesssim \frac{Q(0)}{F(0)}=\frac{1}{2}\frac{\sigma}{\varepsilon}\bigg{|}_{r=0}. \label{mqratio}
\end{equation}
%

We see that, in order to show the occurrence of the shell bounce, we needed somewhat artificial condition of Eq. (\ref{awkcondi2}) 
which is not required in the Einstein case. We may conclude 
that  the shell near the center bounces inevitably for sufficiently small charge in the sense of the condition 
(\ref{mqratio}).

\section{Shell-crossing}

In the previous section, using the kinematical argument, we showed that the shell bounces if the conditions 
(\ref{eincondi}) and (\ref{awkcondi2}) are satisfied for the Einstein, 
Gauss-Bonnet and Lovelock gravity respectively. Hereafter, we assume these conditions.
In this section we will show that the shell-crossing occurs near the center before the 
shell bounces if the charge is relatively small. In order to show this, we shall adopt 
a new coordinate system, called Mass-Area coordinates. This helps us to study on the 
shell crossings as in the four dimensional case of charged dust collapse \cite{Ori:1990,Ori:1991}.

\subsection{Mass-area coordinate}

We first employ the coordinate transformation $(t,r)\rightarrow (F,R)$, where $F$ is the mass function introduced 
in Eq. (\ref{LL6}) and $R$ is the area radius. 
Then, under the above coordinate transformation, the metric transforms as
%
\begin{align}
ds^2&= -A^2dt^2+B^2dr^2+R^2d\Omega_n \notag \\
    &=\bigg[-A^2f_R^2+\left( \frac{B}{R^{\prime}}\right)^2 \left( 1-\dot{R}f_R \right)^2 \bigg] dR^2
     +\bigg[-A^2f_F^2+\left( \frac{B}{R^{\prime}}\right)^2 \dot{R}^2f_F^2  \bigg] dF^2 \notag \\
     &\ \ \ +2\bigg[-A^2f_Rf_F+\left( \frac{B}{R^{\prime}}\right)^2 \left( 1-\dot{R}f_R \right) \dot{R}f_F \bigg] dRdF +R^2d\Omega^2_{n} \notag \\
    &=g_{RR}dR^2+2g_{RF}dRdF+g_{FF}dF^2+R^2d\Omega^2_{n},
\end{align}
%
where $t=f(R,F), r=r(R,F), f_R=\partial_Rf$ and $f_F=\partial_Ff$.
The reversibility of the coordinate transformations gives us
%
\begin{align}
f_R\dot{R}&=1,\label{ctr} \\
r_R&=0,\\
F^{\prime}r_F&=1
\end{align}
and
\begin{equation}
f_FF^{\prime}+f_RR^{\prime}=0, \label{sccondi}
\end{equation}
%
where $r_R=\partial_Rr$ and $r_F=\partial_Fr$. The metric components can be written as
%
\begin{align}
g_{FF}&=-\mathcal{F}^2 \left( u^2-(u^R)^2 \right) ,\\
g_{RF}&=-\mathcal{F}\frac{u}{u^R}
\end{align}
and
\begin{equation}
g_{RR}=-\frac{1}{(u^R)^2} ,
\end{equation}
%
where $u^R=\dot R/A$ is the $R$-component of the shell velocity, and 
$u$ and $\mathcal{F}$ are defined by
%
\begin{eqnarray}
& &u=\frac{R'}{B}= W-\frac{nQQ_N}{2(n-1)R^{n-1}},\label{defu}\\ 
& & \mathcal{F}=\frac{Af_F}{u} \label{defF}.
\end{eqnarray}
%

Using Eq. (\ref{modgeo}), we can show 
%
\begin{equation}
\mathcal{F}_R=-\frac{u^R_F}{u(u^R)^2},\label{FFR}
\end{equation}
%
where ${\cal F}_R = \partial_R {\cal F}$. See Appendix B for the details of the derivation. 

Let us define $\psi$ by 
%
\begin{equation}
\psi =\frac{(u^R)^2+1-u^2}{R^2}. \label{psidef}
\end{equation}
%
Note that $\psi$ satisfies the following equation which is same as Eq. (\ref{basiceq}),
%
\begin{equation}
\sum_{m=1}^kc_m\psi^m =-\frac{nQ^2}{2(n-1)R^{2n}}+\frac{F}{R^{n+1}}. \label{psieq}
\end{equation}
%
Differentiating Eq. (\ref{psidef}) with respect to $F$, $u^R_F$ can be written as 
%
\begin{equation}
u^R_F=\frac{1}{u^R}\Bigl(uu_F+\frac{1}{2}R^2\psi_F \Bigr).
\end{equation}
%
$u_F$ is computed from the definition of $u$ (Eq. (\ref{defu})) as 
%
\begin{equation}
u_F=W_F-\frac{n(Q_N^2+Q_{NN})}{4(n-1)WR^{n-1}},
\end{equation}
%
where we have used $\partial N /\partial F=1/2W$, which is directly derived from Eq. (\ref{FW}). 
Now, we can rewrite $\mathcal{F}_R$ as
%
\begin{equation}
\mathcal{F}_R=-\frac{1}{\left( u^R\right)^3 }\left( W_F-\frac{Q_N^2+QQ_{NN}}{4(n-1)WR^{n-1}}+\frac{R^2\psi_F}{2u}\right) . \label{F_R}
\end{equation}
%
Then the integration of Eq. (\ref{F_R}) gives us
%
\begin{equation}
\mathcal{F}=-\int_r^R dR\frac{1}{2\left( u^R\right)^3 }\left( W_F-\frac{n(Q_N^2+QQ_{NN})}{4(n-1)WR^{n-1}}+\frac{R^2\psi_F}{2u}\right) +\mathcal{H}(F) 
\label{defF},
\end{equation}
%
where $\mathcal{H}$ is the integration function of $F$. 

From Eq. (\ref{sccondi}) we can see that $\mathcal{F}=0$ corresponds to $R^{\prime}=0$ {\it i.e.} 
shell-crossing singularity. In the remaining part, we investigate whether the shell-crossings 
occurs or not for each cases. 

Note that the energy density in the current coordinates is written as 
%
\begin{equation}
\varepsilon =-\frac{n}{4}\frac{1}{4\pi WR^nu^R\mathcal{F}} \label{MAE}.
\end{equation}
%
We observe that $\mathcal{F}$ should be positive in collapsing situations because 
it is assumed that $u^R$ is negative and $\varepsilon $ is positive. 
From now on we consider the marginally bound case ($W=1$) for simplicity.

\subsection{Einstein case}

In the Einstein gravity case, $\psi$ is
%
\begin{equation}
\psi =\frac{F}{R^{n+1}}-\frac{nQ^2}{2(n-1)R^{2n}}
\end{equation}
%
and $\psi_F$ becomes 
%
\begin{equation}
\psi_F=\frac{1}{R^{n+1}}-\frac{nQQ_F}{(n-1)R^{2n}}=\frac{u}{R^{n+1}},
\end{equation}
%
where we used the fact that $Q_F=Q_N/2$ holds for $W=1$(See Eq. (\ref{FW})) and the definition of 
$u$(Eq. (\ref{defu})). 

Then Eq. (\ref{defF}) can be re-written as
%
\begin{equation}
\mathcal{F}=-\int_r^R \frac{dR}{2(u^R)^3R^{n-1}}\left( 1-\frac{n}{2(n-1)}\left( Q_N^2+QQ_{NN}\right) \right) +\mathcal{H}(F) .\label{EinF}
\end{equation}
%
Here we note that $Q_N^2=\mathcal{O}(r^0)$ and $QQ_{NN}=\mathcal{O}(r^2)$ because of $Q=\mathcal{O}(r^{n+1})$ and $N=\mathcal{O}(r^{n+1})$ 
(see Eqs. (\ref{FW}), (\ref{regularity}) and (\ref{regularity2})). Thus, if the condition 
%
\begin{equation}
Q_N^2 =\frac{\sigma^2}{\varepsilon^2}<{2(1-1/n)}
\end{equation}
%
is satisfied, the integrand is positive near the center. 

Since $u^R=\frac{\dot{R}}{A}\sim 0$ near the bounce, $\mathcal{F}$ diverges with the negative sign. 
We see from Eq. (\ref{MAE}) that $\mathcal{F}$ is positive at the initial surface, but $\mathcal{F}$ 
is negative near the shell bounce from the above argument. This implies that the shell crossing singularity 
occurs near the center before the shell bounces if the charged cloud is weakly charged, in the sense of 
$| \frac{\sigma}{\varepsilon }|\leq (2-2/n)^{1/2}$ (Eq. (\ref{eincondi})).

\subsection{Gauss-Bonnet case}

In the Gauss-Bonnet case, $\psi$ is
%
\begin{equation}
\psi =-\frac{1}{2\alpha}\pm \frac{1}{2\alpha}\left( 1+\frac{4\alpha F}{R^{n+1}}-\frac{2\alpha nQ^2}{(n-1)R^{2n}}\right)^{1/2}. \label{gausspsi}
\end{equation}
%
From the condition (\ref{awkcondi2}), $\psi$ is restricted to positive values for the shell near the center
\footnote{First we can see that $QQ_N \simeq r^{n+1}\sigma^2(0)/\epsilon(0)$ is positive near the center ($r=0$). 
And it is easy to see that the function $Z:=1-(1-\frac{nQQ_N}{2(n-1)R^{n-1}})^2$ has only one maximum value 
with respect to $R$, and it tends to zero as $R \to \infty$ and diverges with a negative sign as 
$R \to +0$. In addition, at $t=t_*$, 
$Z$ has a positive value of $1-(1-2Q'F/QF')^2$ because of the condition (\ref{awkcondi2}). 
Since $\psi=\frac{1}{R^2}(\frac{\dot{R}^2}{A^2}+Z)$, it turns out that $\psi$ 
is positive from the beginning to the epoch $t=t_*$ near the center.}. 
Thus we only consider the plus branch of Eq. (\ref{gausspsi}). 

From the direct computation, we can see that 
%
\begin{equation}
\psi_F =\frac{1}{ R^{n+1}\left( 1+\frac{4\alpha F}{R^{n+1}}-\frac{2\alpha nQ^2}{(n-1)R^{2n}}\right)^{1/2}} \left( 1-\frac{n}{2(n-1)}\frac{QQ_N}{R^{n-1}}\right)
\end{equation}
%
and then Eq. (\ref{defF}) becomes 
%
\begin{equation}
\mathcal{F} =-\int_r^R \frac{dR}{2(u^R)^3R^{n-1}}\left( \frac{1}{ \left( 1+\frac{4\alpha F}{R^{n+1}}-\frac{2\alpha nQ^2}{(n-1)R^{2n}}\right)^{1/2} }-\frac{n}{2(n-1)}\left( Q_N^2+QQ_{NN}\right) \right)+\mathcal{H}(F) . \label{GaussF}
\end{equation}
%
We observe that if the condition
%
\begin{equation}
Q_N^2=\frac{\sigma^2}{\varepsilon^2}< \frac{n-1}{n}\frac{2}{ \left( 1+\frac{4\alpha F}{R_*^{n+1}}-\frac{2\alpha nQ^2}{(n-1)R_*^{2n}}\right)^{1/2} }=\frac{n}{n-1}\frac{2}{\left( 1+\frac{2\alpha (n-1)F}{nR_*^{n-1}}\right)^{1/2} }\label{gaussCON}
\end{equation}
%
is satisfied, the integrand of Eq. (\ref{GaussF}) will be positive near the center. 
Here, $R_*$ is defined as $R_*^{n-1}=\frac{n^2}{(n-1)(n+1)}\frac{Q^2}{F}$ 
which maximizes the term $1+\frac{4\alpha F}{R^{n+1}}-\frac{2\alpha nQ^2}{(n-1)R^{2n}} $ 
for given $F$ and $Q$. This leads to the divergence of the integral in  
Eq. (\ref{GaussF}) with the negative sign. We saw from Eq. (\ref{MAE}) that $\mathcal{F}$ 
is positive at the initial surface. On the other hand, $\mathcal{F}$ will be negative 
near the shell bounce. Therefore, we can conclude that the shell near the 
center hits the shell-crossing singularity before the shell bounces 
if the conditions (\ref{awkcondi2}) and (\ref{gaussCON}) are satisfied. 

Now let us examine the meaning of the conditions (\ref{awkcondi2}) and (\ref{gaussCON}) again. 
First, the condition (\ref{gaussCON}) corresponds to
$Q_N^2 <\frac{2(n-1)}{n}$ in the Einstein gravity. This means that the shell is weakly charged compared to mass 
(see Eq. (\ref{eincondi})). Next, under the condition (\ref{gaussCON}), the condition 
(\ref{awkcondi2}) roughly implies that the charge contained in the ball of radius $r$ 
is smaller than the mass. This can be seen from Eq. (\ref{mqratio}). As explained in Sec. IV, we need this condition to show 
the occurrence of shell crossing singularity.

In summary, we can conclude that the shell near the center ($r=0$) 
hits the shell-crossing singularity if the conditions (\ref{awkcondi2}) and (\ref{gaussCON}) 
are satisfied.

\subsection{Lovelock gravity case}

In the Lovelock gravity case, $\mathcal{F}$ is given by
%
\begin{equation}
\mathcal{F} =-\int_r^R \frac{dR}{2(u^R)^3R^{n-1}}\left( \frac{1}{ \Sigma_{m=1}^k mc_m\psi^{m-1} }-\frac{n}{2(n-1)}\left( Q_N^2+QQ_{NN}\right) \right) +\mathcal{H}(F) , \label{LoveF}
\end{equation}
%
and where
%
\begin{equation}
\psi_F =\frac{1}{R^{n+1}\Sigma_{m=1}^kmc_m\psi^{m-1}}\left( 1-\frac{n}{2(n-1)}\frac{QQ_N}{R^{n-1}}\right) .
\end{equation}
%
This is obtained by differentiating Eq. (\ref{eincondi}) with respect to $F$. 
Because of the condition (\ref{awkcondi2}), $\psi$ is restricted to positive values for the 
shell near the center (See the footnote around Eq. (\ref{gausspsi})). As in the Gauss-Bonnet case, if the condition
%
\begin{equation}
Q_N^2=\frac{\sigma^2}{\varepsilon^2}< \min_{R_*\leq R\leq r} \left( \frac{2(n-1)}{n}\frac{1}{ \Sigma_{m=1}^k mc_m\psi^{m-1} } \right) \label{loveCON}
\end{equation}
%
is satisfied, the integrand of Eq. (\ref{LoveF}) will be positive near the center.
From Eq. (\ref{MAE}), $\mathcal{F}$ is positive at the initial surface, but we saw 
that $\mathcal{F}$ will be negative near the shell bounce. Therefore we can conclude 
that the shell near the center hits the shell-crossing singularity if the conditions 
(\ref{awkcondi2}) and (\ref{loveCON}) are satisfied. 

The meaning of the conditions (\ref{awkcondi2}) and (\ref{loveCON}) is the same as in the
Gauss-Bonnet case. The condition (\ref{loveCON}) corresponds to $Q^2_{N}<2(n-1)/n$ in the Einstein case, which means dust is weakly charged as compared to mass. Then the conditions 
(\ref{awkcondi2}) and (\ref{loveCON}) roughly imply that the charge contained in 
the ball of radius $r$ is smaller than the mass. 

\section{Summary and discussion}

In this paper we considered the spherical collapse of a charged dust cloud in the 
Einstein, Gauss-Bonnet and Lovelock gravity. In the case of spherical collapse of 
a dust cloud \cite{Ohashi:2011zz}, all shells of the dust cloud cannot bounce and will 
hit the singularity. But, in the case of collapse of a charged dust cloud, we found 
that the weakly charged shells near the center can bounce kinematically in each 
theories of gravity. This is because of the existence of a repulsive force exerted by the charge. 
We also found that such shells inevitably hit the shell focusing singularity 
in the theories of gravity considered here.

In the higher dimensional Einstein gravity case, we found that the the shells near 
the center satisfying the condition $Q^2_{N}=\left( \frac{\sigma}{\varepsilon}\right)^2 <2(n-1)/n$, where 
$\sigma$ is the electric charge density and $\varepsilon$ is the mass density, can bounce 
{\it kinematically}. We also found that such shells inevitably hit the shell 
crossing singularity. In the four dimensional case, it has been shown that the shells with 
sufficiently small charge inevitably bounce \cite{Ori:1990,Ori:1991}. We found that 
this feature remains to be correct even in the  higher dimensional Einstein gravity.

Next we considered the Gauss-Bonnet and Lovelock gravity cases. In both of the theories, we found that the 
shell near the center can bounce under the condition (\ref{awkcondi2}), which is not imposed in the 
Einstein gravity. We also found that the shell hits the shell-crossing singularity with the condition 
(\ref{awkcondi2}) and (\ref{gaussCON}) in the Gauss-Bonnet case, (\ref{awkcondi2}) and (\ref{loveCON}) 
in the Lovelock gravity case respectively. 
We examined the meaning of the conditions (\ref{awkcondi2}), (\ref{gaussCON}) and (\ref{loveCON}), and 
it turns out 
that these conditions can be interpreted as that the shell is weakly charged compared to mass. Then we can 
conclude that the feature of the shell dynamics in four dimension is shared even in the Lovelock gravity case.

We note that the reason why we imposed the additional condition (\ref{awkcondi2}) in the 
Gauss-Bonnet and Lovelock gravity cases is purely technical. Therefore, our results should be true for more 
general initial data. The condition should be relaxed if we choose the epoch more sophisticated than $t=t_*$ 
which we chose in this paper, or if we perform a numerical analysis. The above remaining issue will be 
addressed in the near future. Since our initial data are specific, it would be interesting to see if there 
exists initial data sets which avoid shell crossing singularities in the Gauss Bonnet and Lovelock gravity theories 
analogous to the four dimensional 
results \cite{Krasinski:2006sb}.

\begin{acknowledgments}
 We thank Tomohiro Takahashi, Kentaro Tanabe, Shunichiro Kinoshita for their useful comments and discussions. 
SO thanks Professor Takashi Nakamura for his continuous encouragement. SO is supported by JSPS Grant-in-Aid for Scientific Research (No. 23-855). TS is supported by Grant-Aid for Scientific Research from MEXT (Nos.~21244033 and 19GS0219). 
This work is also supported in part by MEXT through Grant-in-Aid for the
Global COE Program gThe Next Generation of Physics, Spun from Universality and Emergenceh at Kyoto University.
\end{acknowledgments}

\appendix

\section{Junction between inner and outer solutions}

In this appendix, we will address the junction between the inner and outer solutions.
We assume that the inner solution is
the collapsing dust cloud presented here and the outer solution is the static vacuum black
hole. We follow the argument in Refs. \cite{Maeda:2006pm, Poisson:2004} where
the junction conditions in the Gauss-Bonnet theory and Einstein theory are discussed.

Let $\Sigma $ to be the boundary of the two regions. The metric of the inner solution is
%
\begin{equation}
ds^2=-A^2dt^2+B^2dr^2+R^2d\Omega_n
\end{equation}
%
and that of the outside region is
%
\begin{equation}
ds^2=-f(R)dT^2+\frac{1}{f(R)}dR^2+R^2d\Omega_n . 
\end{equation}
%
We suppose that $\Sigma$ is described by the parametric equations $R=R_{\Sigma}(t)$ and $T=T_{\Sigma}(t)$. 
In addition, it is natural to think that the boundary is comoving, that is, $r=r_0=$constant. 
The induced metric $q$ on $\Sigma$ from the metric of the inner region is written as
%
\begin{equation}
q=-A^2dt^2+R_{\Sigma}^2d\Omega_n .
\end{equation}
%
On the other hand, using the metric of the outer region, it can also be rewritten as
%
\begin{equation}
q=-\left( f(R_{\Sigma})\dot{T}^2_{\Sigma}-\frac{\dot{R_{\Sigma}}^2}{f(R_{\Sigma})}\right) dt^2
+R_{\Sigma}^2d\Omega_n.
\end{equation}
%
Since these should be identical, we see that
%
\begin{equation}
A^2=\left( f(R_{\Sigma})\dot{T}^2_{\Sigma}-\frac{\dot{R_{\Sigma}}^2}{f(R_{\Sigma})}\right)
\end{equation}
%
holds.

The extrinsic curvatures of $\Sigma$ evaluated from the inner metric are
%
\begin{equation}
{}^-K^t_t=-\frac{A^{\prime}}{AB}
\end{equation}
and
\begin{equation}
{}^-K^i_j=\delta^i_j \frac{R^{\prime}_{\Sigma}}{R_{\Sigma}B}.
\end{equation}
%
In terms of the metric of the outer region, the extrinsic curvatures also have another expression 
%
\begin{equation}
{}^{+}K^t_t=\frac{1}{\left( f(R_{\Sigma})\dot{T}^2-\frac{\dot{R}^2_{\Sigma}}{f(R_{\Sigma})}\right)^{3/2}}\left( \dot{R}\ddot{T}-\dot{T}\ddot{R}+\frac{3f^{\prime}}{2f}\dot{R}^2\dot{T}+\frac{\dot{T}^3}{2}ff^{\prime}\right)
\end{equation}
and
\begin{equation}
{}^{+}K^i_j=\delta^i_j \frac{\dot{T}f(R_{\Sigma})}{R_{\Sigma} \left( f(R_{\Sigma})\dot{T}^2-\frac{\dot{R}_{\Sigma}}{f(R_{\Sigma})}\right)^{1/2}}.
\end{equation}
%

The continuity of the extrinsic curvature implies 

%
\begin{equation}
\dot{R}^{\prime}=\frac{A^{\prime}}{A}\dot{R}+\frac{\dot{B}}{B}R^{\prime}
\end{equation}
%
and
%
\begin{equation}
f=-\left( \frac{\dot{R}}{A}\right)^2 +\left( \frac{R^{\prime}}{B} \right)^2 , \label{A11}
\end{equation}
%
where
%
\begin{equation}
f(R_{\Sigma})=1-R^2_{\Sigma}\psi (R_{\Sigma}).
\end{equation}
%
Then, from Eq. (\ref{A11}), we see
%
\begin{equation}
\left( \frac{\dot{R}}{RA}\right)^2 +\frac{1}{R^2}-\left( \frac{R^{\prime}}{RB} \right)^2 =\psi (R_{\Sigma}).\label{newcond}
\end{equation}
%

We recall that $\psi$ satisfies Eq. (\ref{psieq}) 
%
\begin{equation}
\sum_{m=1}^k c_m \psi^{m} =\frac{\mu}{R^{n+1}_{\Sigma}}-\frac{q^2}{R^{2n}_{\Sigma}},
\end{equation}
%
and $\dot{R}$ is the solution to Eq. (\ref{basiceq})
%
\begin{equation}
\sum_{m=1}^k c_m \left( \left( \frac{\dot{R}}{RA}\right)^2 +\frac{1}{R^2}-\left( \frac{R^{\prime}}{RB} \right)^2 \right)^{2m}=\frac{F(r_0)}{R^{n+1}_{\Sigma}}-\frac{Q(r_0)^2}{R^{2n}_{\Sigma}}.
\end{equation}
%
It can be easily checked that Eq. (\ref{newcond}) holds only if
%
\begin{align}
\mu &=F(r_0) \\
q&= Q(r_0).
\end{align}
%
This shows that the inner region can be naturally joined to the outer region if $\mu =F(r_0)$ and $ q=Q(r_0)$ 
is satisfied.

\section{Equation of motion of charged shell in the mass-area coordinate}

In this section, writing down Eq. (\ref{modgeo}) in the mass-area coordinate ($M,R$), we derive 
Eq. (\ref{FFR}). We first note 
that $u^R=\dot R /A$ and the other components of $u^\mu$ vanish. The non-trivial component of Eq. (\ref{modgeo}) 
is 
%
\begin{equation}
\epsilon u^\mu \nabla_\mu u^R=\sigma F^R_{~R}u^R. \label{uur}
\end{equation}
%
Now, since 
%
\begin{equation}
F^R_{~R}=\dot R r_R F^t_{~r}+R' f_R F^r_{~t}=\frac{n}{2}\frac{uQ}{u^R R^n}
\end{equation}
%
and
%
\begin{equation}
\Gamma^R_{RR}=-\frac{u^R_{,R}}{u^R}+\frac{u}{\mathcal{F}u^R}\Bigl( \frac{u^R_{,F}}{(u^R)^3}
+\frac{(\mathcal{F}_{R}u+\mathcal{F}u_{,R})}{u^R} \Bigr),
\end{equation}
%
Eq. (\ref{uur}) gives us 
%
\begin{equation}
\frac{\epsilon u}{\mathcal{F}}\Bigl( 
 \frac{u^R_{,F}}{(u^R)^2}
+(\mathcal{F}_{,R}u+\mathcal{F}u_{,R}) 
\Bigr) = \sigma \frac{n}{2}\frac{uQ}{R^n}.
\end{equation}
%
Using $u_{,R}=nQQ_N/2R^n$(See Eq. (\ref{defu})) and $Q_N=\sigma/\epsilon$(See Eq. (\ref{QNse})), 
we obtain Eq. (\ref{FFR}).



\end{document}